\documentclass[sigconf]{acmart}
\usepackage{url,caption,subcaption,graphicx}
\usepackage{tabularx}

\copyrightyear{2021}
\acmYear{2021}
\setcopyright{acmcopyright}\acmConference[IMC '21]{ACM Internet Measurement Conference}{November 2--4, 2021}{Virtual Event, USA}
\acmBooktitle{ACM Internet Measurement Conference (IMC '21), November 2--4, 2021, Virtual Event, USA}
\acmPrice{15.00}
\acmDOI{10.1145/3487552.3487855}
\acmISBN{978-1-4503-9129-0/21/11}

\usepackage{textcomp}
\usepackage{pgfplots}
\pgfplotsset{width=8.4cm,height=6cm,compat=1.17}
\usepackage{amsmath,xcolor,soul, fontawesome,algpseudocode,algorithm,xspace,multirow,ctable,longtable,colortbl,lscape,pifont}

\sethlcolor{red}
\usepackage[utf8]{inputenc}
\usepackage[font={small}]{caption}
\usepackage{listings}
\usepackage{xcolor}

\definecolor{codegreen}{rgb}{0,0.6,0}
\definecolor{codegray}{rgb}{0.5,0.5,0.5}
\definecolor{codepurple}{rgb}{0.58,0,0.82}
\definecolor{backcolour}{rgb}{0.95,0.95,0.92}

\lstdefinestyle{mystyle}{
  backgroundcolor=\color{backcolour},   commentstyle=\color{codegreen},
  keywordstyle=\color{magenta},
  numberstyle=\tiny\color{codegray},
  stringstyle=\color{codepurple},
  basicstyle=\ttfamily\footnotesize,
  breakatwhitespace=false,         
  breaklines=true,                 
  captionpos=b,                    
  keepspaces=true,                 
  numbers=left,                    
  numbersep=5pt,                  
  showspaces=false,                
  showstringspaces=false,
  showtabs=false,                  
  tabsize=2
}

\usepackage{booktabs,siunitx,comment}

\newcommand{\tool}{{\sc TrackerSift}\xspace}

\begin{document}


\title[TrackerSift: Untangling Mixed Tracking and Functional Web Resources]{TrackerSift: \\Untangling Mixed Tracking and Functional Web Resources}


\author{Abdul Haddi Amjad}
\email{hadiamjad@vt.edu}
\affiliation{%
  \institution{Virginia Tech}
  \country{USA}
}

\author{Danial Saleem}
\email{l174115@lhr.nu.edu}
\affiliation{%
  \institution{FAST-NUCES}
  \country{Pakistan}
}

\author{Muhammad Ali Gulzar}
\email{gulzar@cs.vt.edu}
\affiliation{%
  \institution{Virginia Tech}
  \country{USA}
}

\author{Zubair Shafiq}
\email{zubair@ucdavis.edu}
\affiliation{%
  \institution{University of California, Davis}
  \country{USA}
}

\author{Fareed Zaffar}
\email{fareed.zaffar@lums.edu.pk}
\affiliation{%
  \institution{LUMS}
  \country{Pakistan}
}

\renewcommand{\shortauthors}{Amjad et al.}

\begin{abstract}
%
Trackers have recently started to mix tracking and functional resources to circumvent privacy-enhancing content blocking tools. 
Such mixed web resources put content blockers in a bind: risk breaking legitimate functionality if they act and risk missing privacy-invasive advertising and tracking if they do not. 
In this paper, we propose \tool to progressively classify and untangle mixed web resources (that combine tracking and legitimate functionality) at multiple granularities of analysis (domain, hostname, script, and method).
Using \tool, we conduct a large-scale measurement study of such mixed resources on 100K websites. 
We find that more than 17\% domains, 48\% hostnames, 6\% scripts, and 9\% methods observed in our crawls combine tracking and legitimate functionality. 
While mixed web resources are prevalent across all granularities, \tool is able to attribute 98\% of the script-initiated network requests to either tracking or functional resources at the finest method-level granularity.
Our analysis shows that mixed resources at different granularities are typically served from CDNs or as inlined and bundled scripts, and that blocking them indeed results in breakage of legitimate functionality.
Our results highlight opportunities for finer-grained content blocking to remove mixed resources without breaking legitimate functionality.

\end{abstract}


\maketitle

\section{Introduction}

\textbf{Background \& Motivation.}
Privacy-enhancing content blocking tools such as AdBlock Plus \cite{adblockplus_web}, uBlock Origin \cite{ublockOrigin}, and Brave \cite{brave} are widely used to block online advertising and/or tracking \cite{Garimella17adblocking,Merzdovnik17BlockMeIfYouCanESP,Malloy16AdblockersIMC}. 
Trackers have engaged in the arms race with content blockers via counter-blocking \cite{Nithyanand16ArmsRaceFOCI,Mughees17AntiABPETS} and circumvention \cite{Alrizah19IMCerrorsMisunderstandings,Le21anticvndss}.
In the counter-blocking arms race, trackers attempt to detect users of content blocking tools and give them an ultimatum to disable content blocking.
In the circumvention arms race, trackers attempt to evade filter lists (e.g., EasyList \cite{EasyList}, EasyPrivacy \cite{EasyPrivacy}) used to block ads and trackers, thus rendering content blocking ineffective.
While both arms races persist to date, trackers are increasingly employing circumvention because counter-blocking efforts have not successfully persuaded users to disable content blocking tools \cite{digiday_turnoff,vice_turnoffad,pagefair2017report}.

\noindent \textbf{Limitations of Prior Work.}
Trackers have been using increasingly sophisticated techniques to circumvent content blocking \cite{Alrizah19IMCerrorsMisunderstandings,Le21anticvndss,Bashir:2018:TCC:3278532.3278573}. 
At a high level, circumvention techniques can be classified into two categories. 
One type of circumvention is achieved by frequently changing the network location (e.g., domain or URL) of advertising and tracking resources.
Content blocking tools attempt to address this type of circumvention by updating filter lists promptly and more frequently \cite{Iqbal17AntiABIMC,Sjosten2020filterlists,Vester2018WhoFilterstheFilters,Iqbal20AdGraphSP,siby2021webgraph}. 
The second type of circumvention is achieved by mixing up tracking resources with functional resources, such as serving both from the same network endpoint (e.g., first-party or Content Delivery Network (CDN)) \cite{Alrizah19IMCerrorsMisunderstandings,chen21jssignatures,daocharacterizing}.
Content blocking tools have struggled against this type of circumvention because they are in a no-win situation: they risk breaking legitimate functionality as collateral damage if they act and risk missing privacy-invasive advertising and tracking if they do not.  
While there is anecdotal evidence, the prevalence and modus operandi of this type of circumvention has not been studied in prior literature.

\noindent \textbf{Measurement \& Analysis.}
This paper aims to study the prevalence of mixed resources, which combine tracking and functionality, on the web. 
We present \tool to conduct a large-scale measurement study of mixed resources at different granularities starting from network-level (e.g., domain and hostname) to code-level (e.g., script and method).
\tool's hierarchical analysis sheds light on how tracking and functional web resources can be progressively untangled at increasing levels of finer granularity.
It uses a localization approach to untangle mixed resources beyond the script-level granularity of state-of-the-art content blocking tools.
We show how to classify methods in mixed scripts, which combine tracking and functionality, to localize the code responsible for tracking behavior. 
A key challenge in adapting software fault localization approaches to our problem is to find a rigorous suite of test cases (i.e., inputs labeled with their expected outputs)~\cite{sb1}.
We address this challenge by using filter lists \cite{EasyList,EasyPrivacy} to label tracking and functional behaviors during a web page load.
By pinpointing the genesis of a tracking behavior even when it is mixed with functional behavior (e.g., method in a bundled script), \tool paves the way towards finer-grained content blocking that is more resilient against circumvention than state-of-the-art content blocking tools.

\noindent \textbf{Results.}
Using \tool, our measurements of 100K websites show that 17\% of the 69.3K observed domains are classified as mixed.
%
The requests belonging to mixed domains are served from a total of 26.0K hostnames. 
\tool classifies 48\% of these hostnames as mixed. 
%
The requests belonging to mixed hostnames are served from a total of 350.1K (initiator) scripts. 
\tool classifies 6\% of these scripts as mixed. 
The requests belonging to mixed scripts are initiated from a total of 64.0K script methods.
\tool classifies 9\% of these script methods as mixed.
Our analysis shows that the web resources classified as mixed by \tool are typically served from CDNs or as inlined and bundled scripts, and that blocking them indeed results in breakage of legitimate functionality. 
While mixed web resources are prevalent across all granularities, \tool is able to attribute 98\% of the script-initiated network requests to either tracking or functional resources at the finest method-level granularity.


Our key contributions include:

\begin{itemize}
    \item a \textbf{large-scale measurement and analysis} of the prevalence of mixed web resources; and
    \item a \textbf{hierarchical  localization approach} to untangle mixed web resources.
\end{itemize}

\section{TrackerSift}
\label{sec: method}

 \begin{figure}[!t]
    \centering
        \includegraphics[width=0.49 \textwidth]{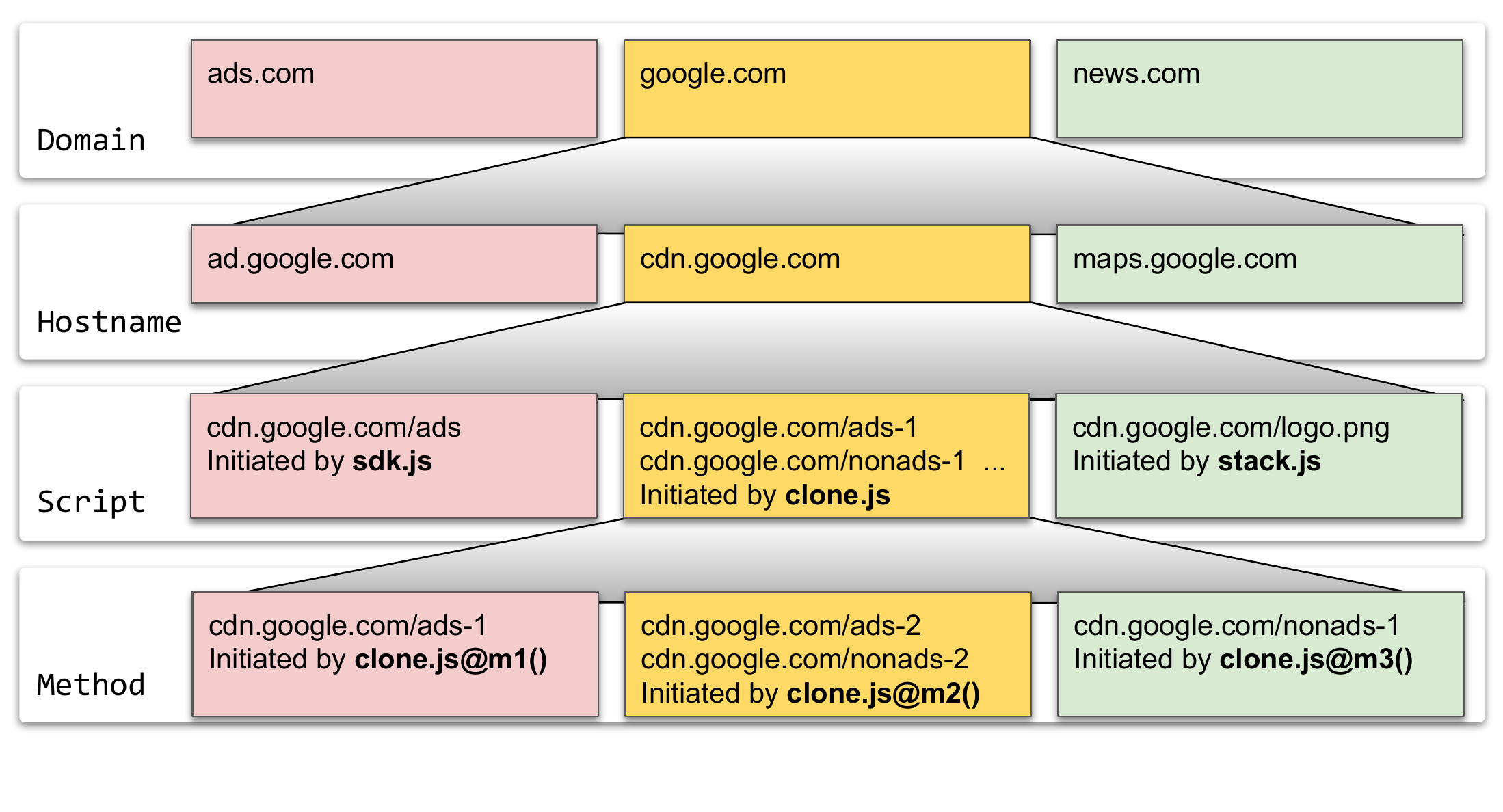}
        \vspace{-0.4in}
        \caption{\tool progressively classifies tracking (red) and functional (green)  resources. For mixed resources (yellow), it proceeds to a finer granularity for further classification.}
        \label{fig:resource-level}
     \vspace{-.2in}
    \end{figure}

In this section, we describe the design of \tool to untangle mixed web resources.  
\tool conducts a hierarchical analysis of web resources to progressively localize tracking resources at increasingly finer granularities if they cannot be separated as functional or tracking at a given granularity. 
\tool needs a test oracle capable of identifying whether a web page's behavior (e.g., network requests) is tracking or functional. 
\tool relies on filter lists, EasyList \cite{EasyList} and EasyPrivacy \cite{EasyPrivacy}, to distinguish between tracking and functional behavior. %
As also illustrated in Figure \ref{fig:resource-level}, we next describe \tool's hierarchical analysis at increasingly finer granularities of domain, hostname, script, and method.

\noindent \textbf{Domain classification.}
As a webpage loads, multiple network requests are typically initiated by scripts on the page to gather content from various network locations addressed by their URLs. We capture such script-initiated requests' URLs and apply filter lists to label them as tracking or functional. We then extract the domain names from request URLs and pass the label from URLs to domain names. 
For each domain, we maintain a tracking count and functional count. All the domains that are classified as tracking or functional are set aside at this level. The rest representing mixed domains serving both tracking and functional requests are further examined at a finer granularity. For instance,  in Figure~\ref{fig:resource-level}, the domain {\tt ads.com} and {\tt news.com} serve solely tracking and solely functional content, respectively. The domain {\tt google.com} serves both and thus needs analysis at a finer granularity.

\begin{figure}[!t]
        \centering
        \includegraphics[width=\columnwidth]{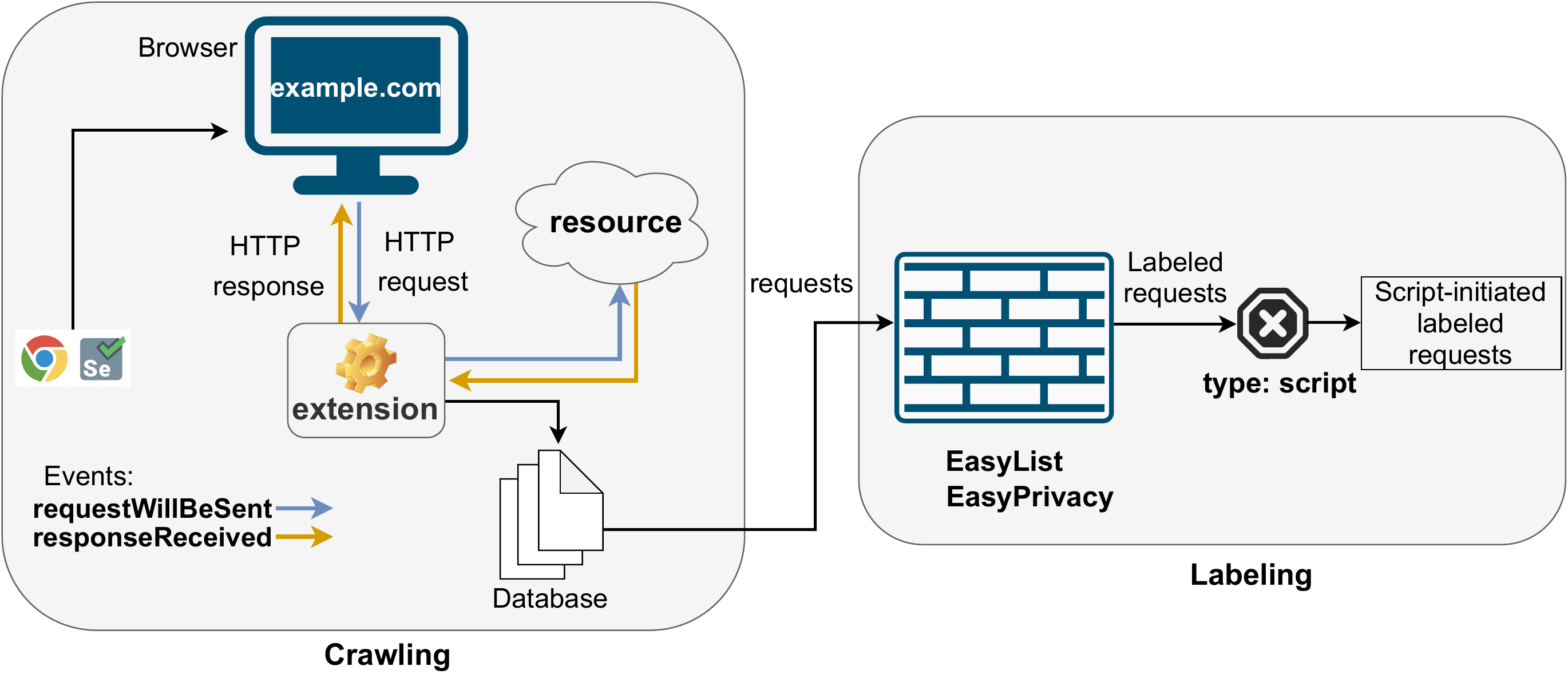}
        \vspace{-.2in}
        \caption{\tool's web crawling and labeling}
        \label{fig:approach}
        \vspace{-.2in}
    \end{figure}

\noindent \textbf{Hostname classification.} At the domain level, we find the requests served by mixed domains and extract their hostnames. We increment the tracking and functional count for each hostname within mixed domains based on the corresponding request's label. The hostnames serving both tracking and functional requests are further analyzed at a finer granularity, while the rest are classified as either tracking or functional. 
In Figure~\ref{fig:resource-level}, {\tt google.com} was previously classified as mixed and therefore, all hostnames belonging to {\tt google.com} need to be examined. We classify {\tt ad.google.com} and  {\tt maps.google.com} as tracking and functional, respectively. In contrast, {\tt cdn.google.com} is mixed and thus needs analysis at a finer granularity.

\noindent \textbf{Script classification.}
We locate the script initiating the request to a mixed hostname and label it as either functional or tracking, reflecting the type of request they initiate. 
Like other levels, we measure the count of tracking and functional requests launched from each script and redistribute those into functional, tracking, and mixed scripts, where mixed scripts will be further analyzed at a finer granularity. 
In Figure~\ref{fig:resource-level}, {\tt sdk.js}, {\tt clone.js}, and {\tt stack.js} all initiate requests to the mixed hostname {\tt cdn.google.com}.
We classify {\tt sdk.js} and {\tt stack.js} as tracking and functional, respectively.
Since {\tt clone.js} requests both tracking and functional resources, it needs analysis at a finer granularity.

\noindent \textbf{Method classification.}
We analyze the corresponding requests for each mixed script and locate the initiator JavaScript methods of each request.  
We then measure the number of tracking and functional requests initiated by each of the isolated methods. 
In the final step, we classify the methods into functional, tracking, and mixed. 
In Figure~\ref{fig:resource-level}, for the mixed script {\tt clone.js}, we classify {\tt m1()} as tracking and {\tt m3()} as functional.
Since {\tt m2()} requests both tracking and functional resources, it is classified as mixed.

  \begin{table}[!t]
    \centering
    \footnotesize
    \caption{Classification of requests at different granularities}
    \vspace{-0.1in}
    \label{table: results1}
    \begin{tabular}{l r r r r r}
\toprule
     {\bf Granularity} & {\bf Tracking} & {\bf Functional} & {\bf Mixed} &  {\bf Separation}  & {\bf Cumulative}\\
     & {\bf (Count)} & {\bf (Count)}  & {\bf (Count)} & {\bf Factor} &  {\bf Separation} \\
     &  &   &  & {\bf (\%)} &   {\bf Factor (\%)}\\
  \midrule
         Domain    & 755,784   &566,810&  1,129,109& 54\% & 54\%\\
        Hostname &   161,604  & 106,542  & 860,963 & 24\% & 65\% \\
         Script &235,157 & 490,295&  135,511 & 84\% & 94\% \\
         Method & 23,819 & 74,223&  37,469 & 72\% & 98\%\\
     \bottomrule
  \end{tabular}
\end{table}

\section{Data}
\label{sec: data}
In this section, we describe \tool's browser instrumentation that crawls websites and labels the collected data.
Note that \tool's hierarchical analysis is post hoc and offline.
Thus, it does not incur any significant overhead during page load other than the browser instrumentation and bookkeeping for labeling.

\noindent \textbf{Crawling.} 
We used Selenium \cite{selenium} with Chrome 79.0.3945.79 to automatically crawl the landing pages of 100K websites that are randomly sampled from the Tranco top-million list \cite{pochat2018tranco} in April 2021. 
Our crawling infrastructure, based on a campus network in North America, comprised of a 13-node cluster with 112 cores at 3.10GHz, 52TB storage, and 832GB memory. 
Each node uses a Docker container to crawl a subset of 100K webpages. 
The average page load time (until \texttt{onLoad} event is fired) for a web page was about 10 seconds. 
Our crawler waits an additional 10 seconds before moving on to the next website. 
Note that the crawling is stateless, i.e., we clear all cookies and other local browser states between consecutive crawls.

As shown in Figure \ref{fig:approach}, our crawler was implemented as a purpose-built Chrome extension that used DevTools \cite{devtools} API to collect the data during crawling.
Specifically, it relies on two \texttt{network} events: \texttt{requestWillBeSent} and \texttt{responseReceived} for capturing relevant information for script-initiated network requests during the page load.
The former event provides detailed information for each HTTP request such as a unique identifier for the request (\texttt{request\_id}), the web page's URL (\texttt{top\_level\_url}), the URL of the document this request is loaded for (\texttt{frame\_\-url}), requested resource type (\texttt{resource\_type}), request header, request timestamp, and a \texttt{call\_stack} object containing the initiator information and the stack trace for script-initiated HTTP requests. 
The latter event provides detailed information for each HTTP response, such as response headers and response body containing the payload.

\noindent \textbf{Labeling.} 
We gather authoritative source labels by applying filter lists to the crawled websites. 
Filter lists are not perfect (e.g., they are slow to update \cite{Sjosten2020filterlists} and are prone to mistakes \cite{Alrizah19IMCerrorsMisunderstandings}) but they are the best available source of labels. 
We use two widely used filter lists that target advertising (EasyList \cite{EasyList}) and tracking (EasyPrivacy \cite{EasyPrivacy}).
These filter lists mainly build of regular expressions that match advertising and/or tracking network requests.
As shown in Figure \ref{fig:approach}, network requests that match EasyList or EasyPrivacy are classified as tracking, otherwise they are classified as functional. 
Note that we maintain the call stack that contains the ancestral scripts that in turn triggered a script-initiated network request (e.g., \texttt{XMLHTTPRequest} fetches). 
For asynchronous JavaScript, the stack track that preceded the request is prepended in the stack. 
Thus, for script-initiated network requests, we ensure that if a request is classified as tracking or functional, its ancestral scripts in the stack are also classified as such.
Since network requests that are not script-initiated can not be trivially classified as tracking or functional, we exclude them from our analysis.

   \begin{table}[!t]
    \centering
    \footnotesize
    \caption{Classification of resources at different granularities}
      \vspace{-0.1in}
    \label{table: results}
    \begin{tabular}{l r r r r}
     \toprule
     {\bf Granularity} & {\bf Tracking} & {\bf Functional} & {\bf Mixed} & {\bf Separation}\\ 
     & {\bf (Count)} & {\bf (Count)}  & {\bf (Count)} & {\bf Factor} \\
     & & & & {\bf (\%)}\\
     \midrule
         Domain    & 6,493   & 50,938 &  11,861 & 83\%\\
         Hostname &   4,429  & 9,248   & 12,383 &52\%\\
         Script & 194,156 & 134,726 &  21,168 & 94\% \\
         Method & 17,940 & 40,500 &  5,579 & 91\%\\
     \bottomrule
  \end{tabular}
\end{table}

\section{Results}
\label{sec: results}

\begin{figure*}[!t]
\begin{subfigure}{.24\textwidth}
    \includegraphics[width=.99\linewidth]{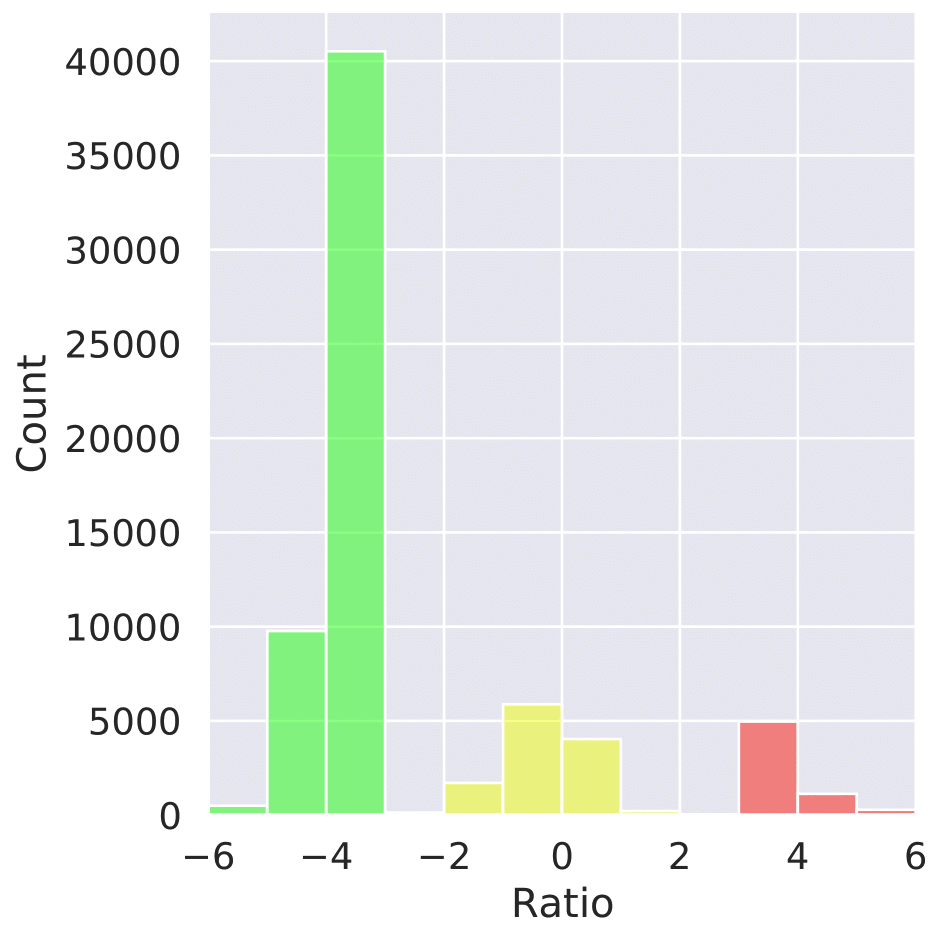}
    \caption{domain}
    \label{fig:domain-analysis}
\end{subfigure}
\begin{subfigure}{.24\textwidth}
    \includegraphics[width=.99\linewidth]{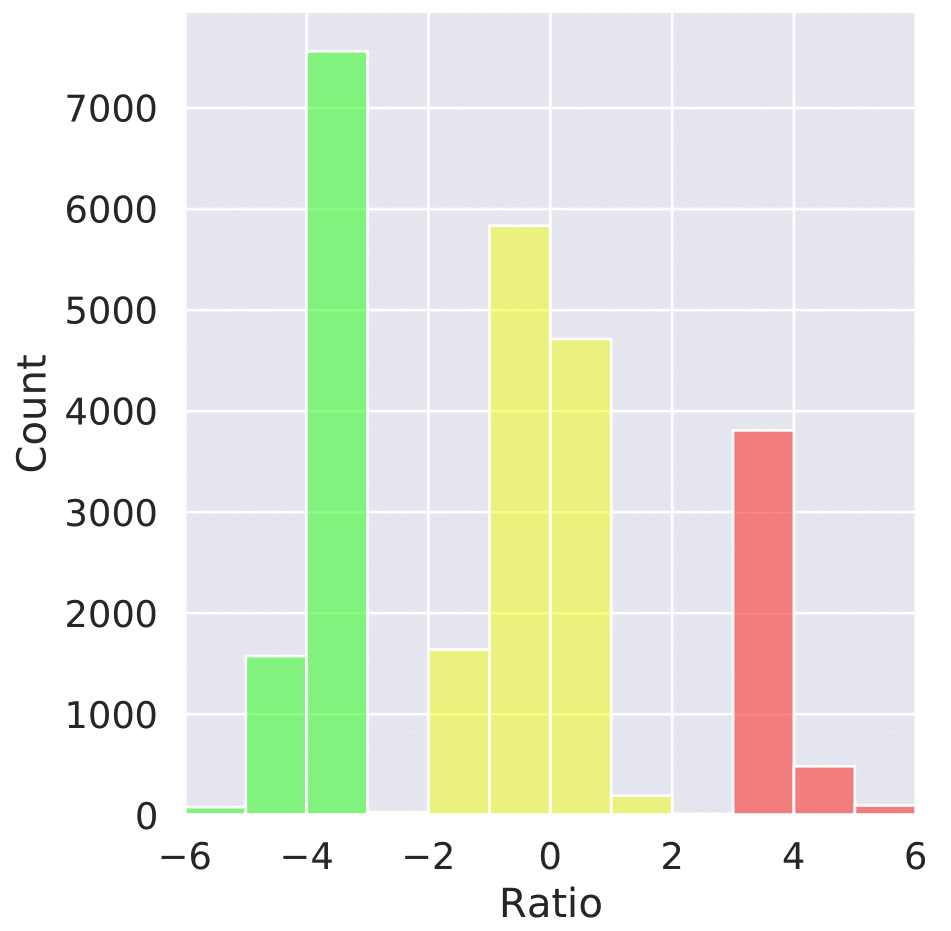}
    \caption{hostname}
    \label{fig:hostname-analysis}
\end{subfigure}
\begin{subfigure}{.24\textwidth}
    \includegraphics[width=.99\linewidth]{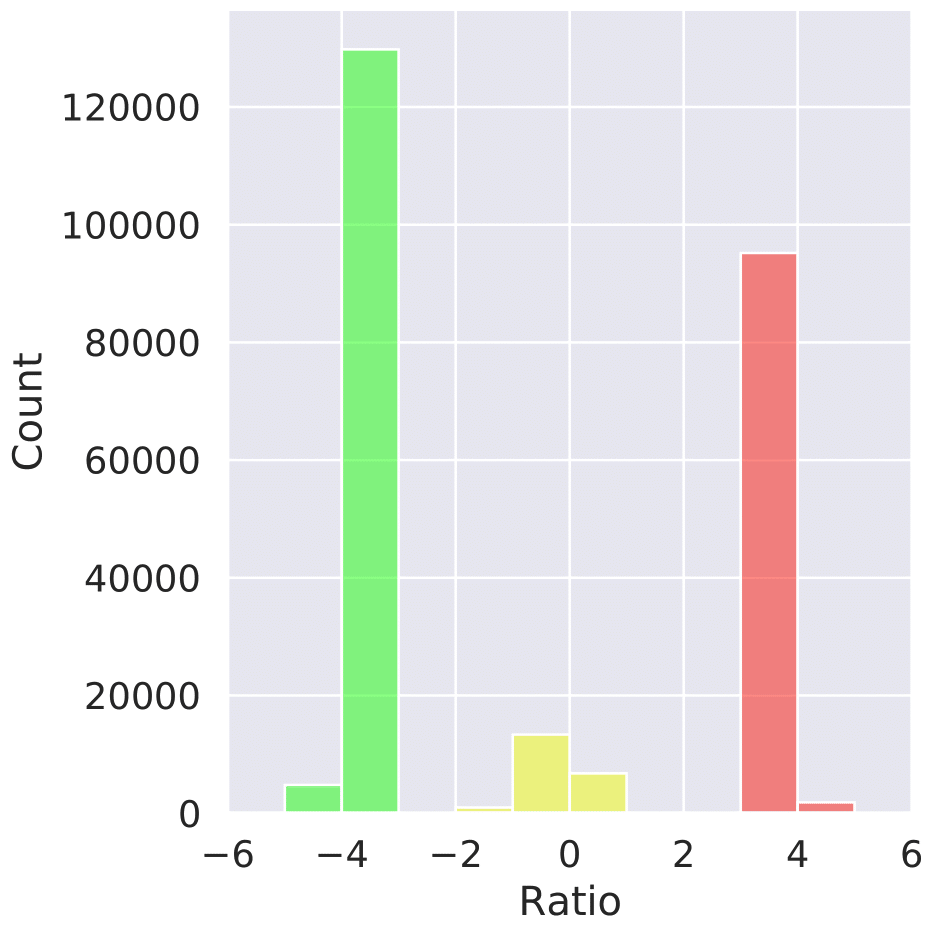}
    \caption{script URL}
    \label{fig:script-analysis}
\end{subfigure}
\begin{subfigure}{.24\textwidth}
    \includegraphics[width=.99\linewidth]{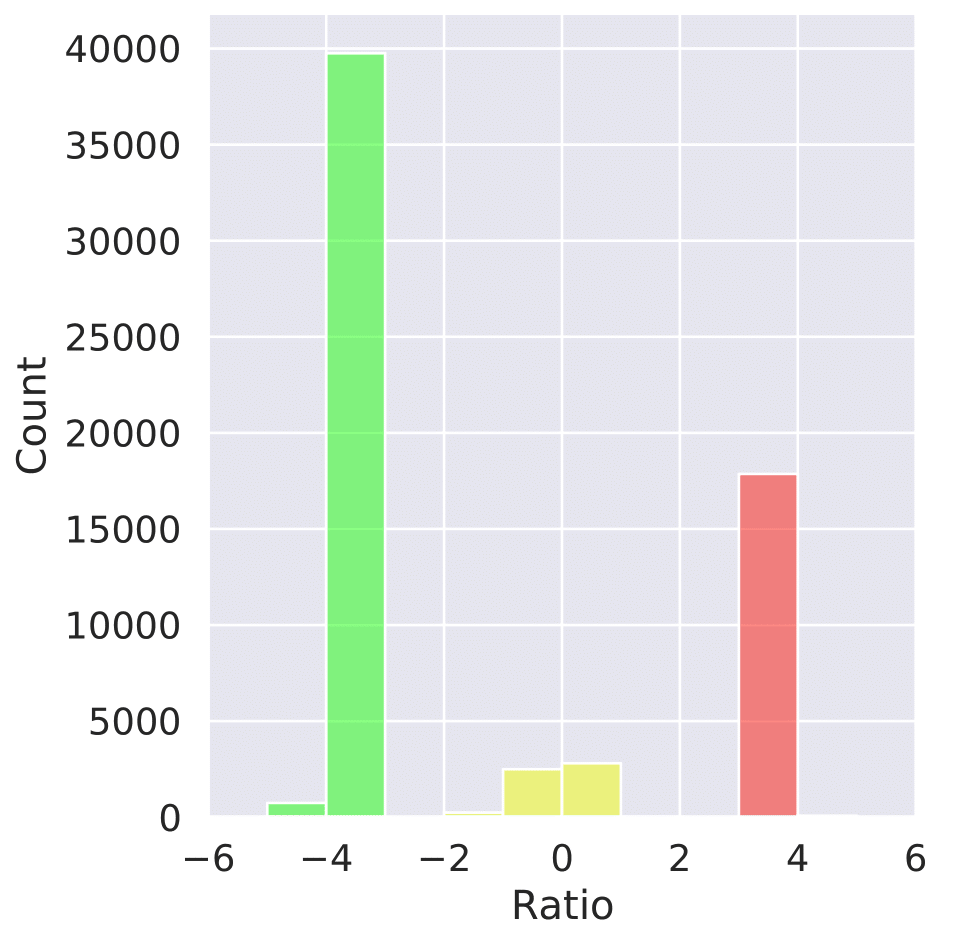}
    \vspace{-.1in}
    \caption{script method}
    \label{fig:scriptmethod-analysis}
    \end{subfigure}
\vspace{-.08in}
\caption{Distribution of resources at increasingly finer granularities.  Y-axis shows the count of unique (a) domains, (b) hostnames, (c) scripts, and (d) script methods. X-axis represents the common logarithmic ratio of the number of tracking to functional requests. Interval (-$\infty$,-2] is classified as functional (green), (-2,2) is classified as mixed (yellow), and [2,$\infty$) is classified as tracking (red).} 
\vspace{-0.1in}
\label{fig:histograms}
\end{figure*}

\noindent \textbf{Classifying Mixed Resources.}
We compute the logarithmic ratio of the number of tracking to functional network requests to quantify the mixing of tracking and functional resources. 

\begin{equation}
    ratio =\log\left(\frac{\#~of~tracking~requests}{\#~of~functional~requests}\right)
    \label{equation: ratio}
\end{equation}
At each granularity, we classify resources with the common logarithmic ratio less than -2 as functional because they triggered 100$\times$ more functional requests than tracking requests. 
Similarly, we classify resources with the common logarithmic ratio of more than 2 as tracking because they triggered 100$\times$ more tracking requests than functional requests. 
The resources with the common logarithmic ratio between -2 and 2 are classified as mixed.
We analyze the suitability of the selected classification threshold using sensitivity analysis later in Section \ref{sec: discussion}.

\noindent \textbf{Results Summary.}
{Table \ref{table: results1}} summarizes the results of our crawls of the landing pages of 100K websites. 
Using the aforementioned classification, we are able to attribute 54\% of the 2.43 million script-initiated network requests to tracking or functional domains. 
The remaining 46\% (1129K) of the 2.43 million requests attribute to mixed domains that are further analyzed at the hostname-level.
We are able to attribute 24\% of the requests from mixed domains to tracking or functional hostnames. 
The remaining 76\% (860K) of the requests attribute to mixed hostnames that are further analyzed at the script URL-level.
We are able to attribute 84\% of the requests from mixed hostnames to tracking or functional script URLs. 
The remaining 16\% (135K) of the requests attribute to mixed script URLs that are further analyzed at the script method-level.
We are able to attribute 72\% of the requests from mixed script URLs to tracking or functional script methods. 
This leaves us with less than 2\% (37K) requests that cannot be attributed by \tool to tracking or functional web resources and require further analysis.

Next, we analyze the distribution of the ratio of tracking to functional requests by web resources at different granularities of domain, hostname, script URL, and script method in Figure \ref{fig:histograms}. 
Table \ref{table: results} shows the breakdown of web resources classified as tracking, functional, and mixed at different granularities.

\noindent \textbf{Domain classification.}
2451K requests in our dataset are served from a total of 69,292 domains (eTLD+1).
Figure \ref{fig:domain-analysis} shows three distinct peaks: [2, $\infty$) serve tracking requests, (-$\infty$, -2] serve functional requests, and (-2, 2) serve both tracking and functional requests. 
We can filter 31\% of the requests by classifying 6,493 domains that lie in the [2, $\infty$) interval as tracking.
Notable tracking domains include  \url{google-analytics.com}, \url{doubleclick.net}, and \url{googleadservices.com}, \url{bing.com}.
We can filter 23\% of the requests by classifying 50,938 domains that lie in the (-$\infty$, -2] interval as functional.
Notable functional domains include CDN and other content hosting domains \url{twimg.com},  \url{zychr.com}, \url{fbcdn.ne}, \url{w.org}, and \url{parastorage.com}.
However, 46\% of requests are served by 11,861 mixed domains that lie in the (-2, 2) interval.
These mixed domains cannot be safely filtered due to the risk of breaking legitimate functionality, and not filtering them results in allowing tracking. 
Notable mixed domains include \url{gstatic.com},  \url{google.com}, \url{facebook.com}, \url{facebook.net}, and \url{wp.com}.

\noindent \textbf{Hostname classification.}
1129K requests belonging to mixed domains are served from a total of 26,060 hostnames.
Figure \ref{fig:hostname-analysis} shows three distinct peaks representing hostnames that {serve tracking}, functional, or both tracking and functional requests.
We can filter 14\% of the requests by classifying 4,429 hostnames that lie in the [2, $\infty$) interval as tracking.
We can filter 9\% of the requests by classifying 9,248 hostnames that lie in the (-$\infty$, -2] interval as functional.
However, 76\% of the requests are served by 12,383 hostnames that lie in the (-2, 2) interval are classified as mixed.
Again, these mixed hostnames cannot be safely filtered due to the risk of breaking legitimate functionality, and not filtering them results in allowing tracking. 
Take the example of hostnames of a popular mixed domain \url{wp.com}. 
The requests from \url{wp.com} are served from tracking hostnames such as \url{pixel.wp.com} and \url{stats.wp.com}, functional hostnames such as \url{widgets.wp.com} and \url{c0.wp.com}, and mixed hostnames such as \url{i0.wp.com} and \url{i1.wp.com}.

\noindent \textbf{Script classification.}
860K requests belonging to mixed hostnames are served from a total of 350,050 initiator scripts.
Figure \ref{fig:script-analysis} again shows three distinct peaks representing scripts that serve tracking, functional, or both tracking and functional requests.
We can filter 27\% of the requests by classifying 194,156 scripts that lie in the [2, $\infty$) interval as tracking.
We can filter 57\% of the requests by classifying 134,726 scripts that lie in the (-$\infty$, -2] interval as functional.
However, 16\% of the requests are served by 21,168 scripts that lie in the (-2, 2) interval are classified as mixed.
These mixed scripts cannot be safely filtered due to the risk of breaking legitimate functionality, and not filtering them results in allowing tracking. 
For example, let's analyze the initiator scripts of a mixed hostname \url{i1.wp.com}. 
The requests to this hostname are the result of different initiator scripts on the webpage \url{www.ibn24.tv}.
Specifically, a tracking request to \url{i1.wp.com} is initiated by the script \url{show\_ads\_impl\_fy2019.js} and a functional request to \url{i1.wp.com} is initiated by the script \url{jquery.min.js}. 
As another example, on the webpage \url{somosinvictos.com}, both tracking and functional requests to \url{i1.wp.com} are initiated by the mixed script \url{lazysizes.min.js}.
Note that the scripts classified as tracking initiate requests to well-known advertising and tracking domains.
For example, script \url{uc.js} served by \url{consent.cookiebot.com} initiated requests to googleadservices.com, doubleclick.net, and amazonadsystem.com.


\noindent \textbf{Method classification.}
135K requests belonging to mixed scripts are served from a total of 64,019 script methods.
Figure \ref{fig:scriptmethod-analysis} again shows three distinct peaks representing methods that serve tracking, functional, or both tracking and functional requests.
We can filter 17\% of the requests by classifying 17,940 methods that lie in the [2, $\infty$) interval as tracking.
We can filter 55\% of the requests by classifying 40,500 methods that lie in the (-$\infty$, -2] interval as functional.
However, 28\% of the requests are served by 5,579 methods that lie in the (-2, 2) interval are classified as mixed.
These mixed methods cannot be safely filtered due to the risk of breaking legitimate functionality, and not filtering them results in allowing tracking. 
For example, let's analyze script methods for a  mixed script \url{tfa.js} on the webpage \url{hubblecontacts.com}. 
While both tracking and functional requests are initiated by the script, the tracking request was initiated by \textit{get} method, and the functional request was initiated by \textit{X} method.
As another example, let's analyze script methods for a mixed script \url{app.js} on the webpage \url{radioshack.com.mx}. 
In this case, both tracking and functional requests are initiated by the mixed script method \textit{Pa.xhrRequest}.

\section{Discussion}
\label{sec: discussion}
In this section, we discuss some case studies, opportunities for future work, and limitations.

\noindent \textbf{Circumvention strategies.}
There are two common techniques for mixing tracking and functional resources.
\par{(1) \textit{Script inlining:} Despite potential security risks, publishers are willing to inline external JavaScript code snippets (as opposed to including external scripts using the \texttt{src} attribute) for performance reasons as well as for circumvention \cite{Nikiforakis12jsinclusion,Lauinger17JSlibraries}.
For example, we find that the Facebook pixel \cite{pixel} is inlined on a large number of websites to assist with targeting Facebook ad campaigns and conversion tracking. }
\par{(2) \textit{Script Bundling:} 
Publishers also bundle multiple external scripts from different organizations with intertwined dependencies for simplicity and performance reasons. 
JavaScript bundlers, such as webpack \cite{webpack} and browserify  \cite{browserify}, use dependency analysis to bundle multiple scripts into one or a handful of bundled scripts. 
For example, \url{pressl.co} serves a script  \url{app.9115af433836fd824ec7.js} that is bundled using the webpack \cite{webpack}.
This bundled script includes the aforementioned Facebook pixel and code to load functional resources from a first-party hostname.
Existing content blocking tools struggle to block inlined and bundled tracking scripts without the risk of breaking legitimate site functionality.
Finer-grained detection by \tool presents an opportunity to handle such scripts by localizing the methods that implement tracking.}

\begin{figure}
\begin{tikzpicture}
\begin{axis}[
    title={},
    height = 0.5\linewidth,
    xlabel={Classification Threshold},
    ylabel={\% Mixed Scripts},
    xmin=1, xmax=3.0,
    ymin=5.70, ymax=6.10,
    xtick={1.0,1.1,1.2,1.3,1.4,1.5,1.6,1.7,1.8,1.9,2,2.1,2.2,2.3,2.4,2.5,2.6,2.7,2.8,2.9,3.0},
    ytick={5.70,5.75,5.80,5.85,5.90,5.95,6.00,6.05,6.10},
    legend pos=north west,
    ymajorgrids=true,
    grid style=dashed,
    x tick label style={rotate=90,anchor=east}
]

\addplot[
    color=blue,
    mark=x,
    ]
    coordinates {
    (1,5.745)(1.1,5.872)(1.2,5.935)(1.3,5.973)(1.4,6.004)(1.5,6.018)(1.6,6.030)(1.7,6.038)(1.8,6.043)(1.9,6.045)(2,6.046)(2.1,6.047)(2.2,6.047)(2.3,6.048)(2.4,6.048)(2.5,6.049)(2.6,6.049)(2.7,6.059)(2.8,6.068)(2.9,6.090)(3,6.095)
    };
\end{axis}
\end{tikzpicture}

\vspace{-2.5ex}
\caption{Sensitivity analysis of the classification threshold (default is -2 and 2) by studying the proportion of mixed scripts as a function of varying thresholds. The X-axis represents the threshold buckets. For example, 1.5 represents (-1.5, 1.5).}
\label{fig:senstivity}
\vspace{-4.5ex}
\end{figure}
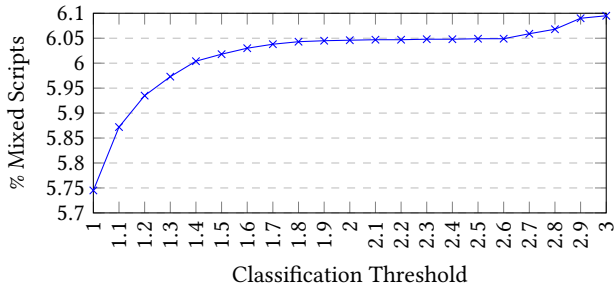


\noindent \textbf{Threshold sensitivity analysis.}
We set the classification threshold to a symmetric value of (-2,2) for classifying mixed resources in Equation \ref{equation: ratio}.
To assess our choice of the threshold, we analyze the sensitivity of script classification results in Figure \ref{fig:senstivity}.
Similar trends are observed for domain, hostname, and method classification. 
The plot shows the percentage of scripts classified as mixed as we vary the threshold from 1 to 3 in increments of 0.1 
Note that the curve plateaus around our selected threshold of 2.
Thus, we conclude that our choice of the threshold is stable and reasonably separates mixed resources from tracking and functional resources.

\noindent \textbf{Breakage analysis.}
We conducted manual analysis to assess whether blocking mixed resources results in breakage of legitimate functionality. 
To assess functionality breakage, we load a random sample of websites with (treatment) and without (control) blocking mixed scripts as classified by \tool.
We label breakage as: \textit{major} if the core functionality such as search bar, menu, images, and page navigation is broken in treatment but not in control; \textit{minor:} if the secondary functionality such as comment/review sections, media widgets, video player, and icons is broken in treatment but not in control; and \textit{none:} if the core and secondary functionalities of the website are same in treatment and control. Note that we consider missing ads as no breakage.
Table \ref{table: breakage} shows our breakage analysis on a representative sample of 10 websites. 
We note major or minor breakage in all except one case.
Thus, we conclude that mixed web resources indeed cannot be safely blocked by existing content blocking tools. 

\begin{table}[!t]
    \centering
    \footnotesize
    \caption{Manual analysis of breakage caused by blocking mixed scripts on randomly selected 10 websites.}
    \vspace{-0.1in}
    \label{table: breakage}
\begin{tabularx}{8.4 cm}{ X l c X}
\toprule
\textbf{Website}         & \textbf{Mixed Script} & \textbf{Breakage} & \textbf{Comment} \\
\midrule
caremanagem\-entmatters.co\-.uk& \url{jquery.min.js} & Minor & scroll bar and two widgets missing \\
\hline
gratis.com& \url{main.js} & Major & page did not load \\
\hline
forevernew\-.com.au& \url{require.js} & Major & multiple page banners missing \\
\hline
flamesnation\-.ca& \url{player.js} & Minor & video pop missing \\
\hline
biba.in& {MJ\_Static-Built.js} & Major & page did not load \\
\hline
ecomarket.ru& {2.0c9c64b2.\-chunk.js} & Major & page did not load \\
\hline
peachjohn.co\-.jp& {jquery-1.11.2.\-min.js} & Major & navigation and scroll bar missing \\
\hline
shoobs.com& \url{widgets.js} & None & no visible functionality breakage \\
\hline
editorajusp\-odivm.com.br& \url{jquery.js} & Major & navigation and scroll bar missing \\
\hline
resourceworld\-.com& \url{jquery.min.js} & Major &  navigation bar and images missing \\
\hline
\end{tabularx}
\vspace{-3.5ex}
\end{table}

\noindent \textbf{Blocking mixed scripts.} 
When \tool classifies a mixed script with different tracking and functional methods, we can simply remove tracking methods to generate a surrogate script that can then be used to shim the mixed script at runtime. 
Existing content blockers such as NoScript, uBlock Origin, AdGuard, and Firefox SmartBlock use surrogate scripts to block tracking by mixed scripts while avoiding breakage \cite{noscriptsurrogates,ubosurrogates,adguardsurrogates,mozilla}. 
However, these surrogate scripts are currently manually designed \cite{security}.
\tool can help scale up the process of generating surrogate scripts by automatically detecting and removing tracking methods in mixed scripts. 
Note that removing tracking methods is tricky because simply removing them risks functionality breakage due to potential coverage issues of dynamic analysis. 
To mitigate this concern, we plan to explore a more conservative approach using a guard\textemdash a predicate that blocks  tracking execution but allows functional execution. 
Such a predicate has a similar structure to that of an {\tt assertion}. 
We envision using classic invariant inference techniques~\cite{saswat2016pie, daikon} on a tracking method's calling context, scope, and arguments to generate a program invariant that holds across all tracking invocations. 
If an online invocation satisfies the invariant, the guard will block the execution.
A key challenge in this approach is collecting the context information, e.g.,  program scope, method arguments, and stack trace, for each request initiated by the mixed method at runtime.
We plan to address these challenges in leveraging \tool for generating safe surrogate scripts in our future work.

    \begin{figure*}[!t]
    \centering
      \includegraphics[width=.90\textwidth]{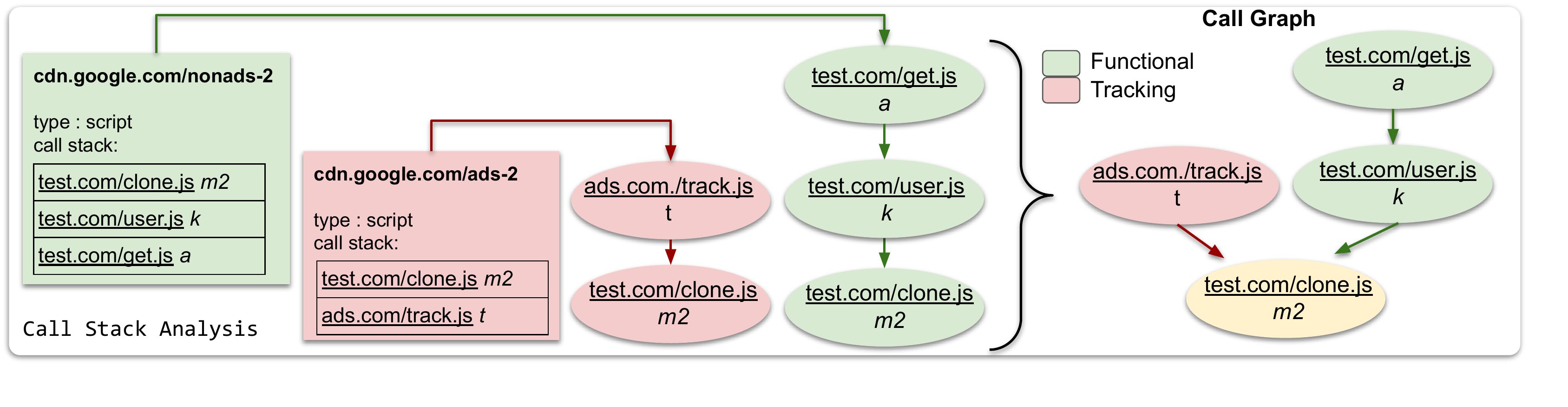}
        \vspace{-.35in}
        \caption{Call stack analysis for the requests \textit{ads-2 }and \textit{nonads-2} that can not be separated at method level i.e. \textit{m2}.  Call stack is analyzed to identify the first point of divergence i.e \textit{track.js t} and it could be removed to block the tracking request. }
        \label{fig:analysis}
        \vspace{-0.1in}
    \end{figure*}

\noindent \textbf{Blocking mixed methods.} 
Our analysis shows that \tool's separation factor is 91\% even at the finest granularity. 
This leaves 5.6K mixed methods that cannot be safely blocked. 
One possible direction is to apply \tool in the \emph{context} of a mixed method initiating a request. 
We can define \emph{context} as calling context, program scope, or parameters to the mixed method. 
In the case of calling context, we can perform a call stack analysis that takes a snapshot of a mixed method's stack trace when the method initiates a tracking or functional request. 
We hope to see distinct stack traces from tracking and functional requests by a mixed method. 
We can consolidate the stack traces of a mixed method and locate the point of divergence, i.e., a method in the stack trace that only participates in tracking requests. 
We hypothesize that removing such a method will break the chain of methods needed to invoke a tracking behavior, thus removing the tracking behavior.

Figure~\ref{fig:analysis} illustrates our proposed call stack analysis. 
It shows the snapshot of stack traces of requests {\tt nonads-2} and {\tt ads-2}. These requests are initiated by a mixed method {\tt m2()} on the webpage. 
The two stack traces are merged to form a call graph where each node represents a unique script and method, and an edge represents a caller-callee relationship. 
The yellow color indicates that a node participates in invoking both tracking and functional requests. {\tt t} in \url{track.js} is the point of divergence since it only participates in the tracking trace. 
Therefore, {\tt t} is most likely to originate a tracking behavior which makes it a good candidate for removal.



\noindent \textbf{Limitations.}
We briefly acknowledge a few limitations our measurement and analysis.
First, our web crawls do not provide full coverage of the events triggered by user interactions (e.g., scroll, click). 
This is a general limitation of dynamic analysis and can be mitigated by using a forced execution framework to execute other possible paths \cite{kim2017j}. 
Second, our method-level analysis does not distinguish between different anonymous functions in a script and treats them as part of the same method. 
This limitation can be addressed by using the line and column number information available for each method invocation in the call stack.  
Finally, our web crawls are limited to the landing pages and the results might vary for internal pages \cite{aqeel2020landing_internal_pages_hispar}.
As part of our future work, we plan to deploy \tool on internal pages as well.

%
%

\section{Related Work}
\label{sec: related work}
We summarize closely related work documenting anecdotal evidence of circumvention by mixing up tracking and functional resources.
Most notably, Alrizah et al. \cite{Alrizah19IMCerrorsMisunderstandings} and Chen et al. \cite{chen21jssignatures} showed how first-party hosting and script inlining or bundling is being used by trackers to circumvent filter lists used by content blockers.
Alrizah et al. \cite{Alrizah19IMCerrorsMisunderstandings} documented a variety of attacks on content blocking tools, including both counter-blocking and circumvention attacks. 
Among other things, they showed that some websites circumvent filter lists by mixing tracking and functional resources through techniques such as script inlining. 
These websites essentially have a ``self-defacement'' strategy, where content blockers risk breaking legitimate functionality as collateral damage if they act and risk missing privacy-invasive advertising and tracking if they do not. 
Chen et al. \cite{chen21jssignatures} leveraged their JavaScript signature approach to document about 500 false negative cases where tracking scripts were inlined or bundled for successful circumvention. 
Relatedly, trackers have started to exploit techniques such as CDN proxies (i.e., serve functional and tracking resources from the same CDN server) \cite{Le21anticvndss} and CNAME cloaking (i.e., masquerade third-party tracking resources from first-party using a minor change in DNS records) \cite{Medium_CNAME,daocharacterizing} to assist with implementing these circumvention techniques.

The problem of localizing tracking-inducing code shares similarities with prior research on fault-inducing code localization.
For example, spectra-based fault localization (SBFL) \cite{sb1,sb2,Eric,agarwal,pearson,Souza} collect statement coverage profiles of each test, passing or failing, to localize the lines of code that are most likely to induce a test failure.
Bela et al. \cite{call} and Laghari et al. \cite{laghari} presented a call frequency-based SBFL technique. 
Instead of coverage information, they use the frequency of method occurrence in the call stack of failing test cases. 
A method that appears more in the failing call stack of failing test cases is more likely to be faulty. 
In \tool, methods responsible for more frequently initiating tracking requests than functional requests is classified as  tracking. 
Abreu et al. \cite{Zoe} studied how accurate these SBFL techniques are, and their accuracy is independent of the quality of test design. 
Jiang et al. \cite{Ji} used call stack to localize the null pointer exception, and Gong et al. \cite{Go} generated call stack traces to successfully identify 65\% of the root cause of the crashing faults. 
One common limitation across most fault-localization approaches is that they require an extensive test suite capable of exercising faulty behavior, along with an instrumented runtime to collect statement-level coverage. 
\tool overcomes these limitations by using filter lists as test oracle during page load time and uses an instrumented browser to capture fine-grained coverage. 

\vspace{.005in}
\section{Conclusion}
\label{sec: conclusion}
We presented \tool, a hierarchical approach to progressively untangle mixed resources at increasing levels of finer granularity from network-level (e.g., domain and hostname) to code-level (e.g., script and method).
We deployed \tool on 100K websites to study the prevalence of mixed web resources across different granularities. 
\tool classified more than 17\% domains, 48\% hostnames, 6\% scripts, and 9\% methods as mixed.
Overall, \tool was able to attribute 98\% of all requests to tracking or functional resources by the finest level of granularity.
Our results highlighted opportunities for finer-grained content blocking to remove mixed resources without breaking legitimate site functionality.
\tool can be used to automatically generate surrogate scripts to shim mixed web resources. 


\section*{Acknowledgements}
This work is supported in part by the National Science Foundation under grant numbers 2051592, 2102347, 2103038, 2103439, and 2106420.
We would like to thank our shepherd, Paul Barford, and the anonymous IMC reviewers, for their constructive feedback. 
We would also like to thank Haris Amjad for his valuable input to help improve the quality of visualizations in the paper.

\bibliographystyle{ACM-Reference-Format}
\bibliography{sample-base}

\end{document}